\documentclass[%
superscriptaddress,
nofootinbib,
nobibnotes,
amsmath,amssymb,
aps,
floatfix,
]{revtex4-2}

\usepackage{placeins}
\usepackage{booktabs}
\usepackage{dcolumn}
\usepackage{array}
\usepackage{longtable}
\usepackage{float}
\usepackage{hyphenat}
\usepackage{natbib}
\usepackage{hyperref}

\usepackage{amsmath,amsfonts,amsthm,amssymb}
\usepackage{bm,graphicx,graphics,color}
\usepackage{epsf,epsfig}
\usepackage[all]{xy}
\newcommand*{\B}[1]{\ifmmode\bm{#1}\else\textbf{#1}\fi}

\def\bx{\mathbf{x}}

\def\bv{\mathbf{v}}

\def\bp{\mathbf{p}}

\def\no{\nonumber}
\def\lb{\label}
\def\be{\begin{equation}}
\def\ee#1{\label{#1}\end{equation}}
\newcommand{\ben}{\begin{eqnarray}}
\newcommand{\een}{\end{eqnarray}}

\begin{document}

\title{Post-Newtonian Kinetic Theory}

\author{Gilberto M. Kremer}
\email{kremer@fisica.ufpr.br}
\affiliation{Departamento de F\'{i}sica, Universidade Federal do Paran\'{a}, Curitiba 81531-980, Brazil}

\begin{abstract}
A kinetic theory for relativistic gases in the presence of gravitational fields is developed in the second post-Newtonian approximation. The corresponding  Boltzmann equation is determined from the evolution  of the one-particle distribution function with respect to the proper time along the world line of the particle. From the knowledge of the equilibrium Maxwell-J\"uttner distribution function in the second post-Newtonian approximation the components of the particle four-flow and energy-momentum tensor are obtained. The Eulerian hydrodynamic equations for the mass density, mass-energy density and momentum density in the second post-Newtonian approximation are determined from the Boltzmann equation. It is shown that the combination of the hydrodynamic equations of mass and mass-energy densities leads to the  hydrodynamic equation for  the internal energy density in the first post-Newtonian approximation. 
\end{abstract}
\maketitle 

\section{Introduction}
\lb{s.1}
Post-Newtonian theory is a method of successive approximations in powers of $1/c^2$ for the solution of Einstein's field equations which was proposed  by Einstein, Infeld and Hoffmann \cite{Eins} in 1938. In this method the components of the metric tensor in the order $\mathcal{O}(c^{n})$ which appear in the Ricci tensor of Einstein's field equations are determined  from the knowledge of the energy-momentum tensor in the order $\mathcal{O}(c^{n-2})$. The first post-Newtonian hydrodynamic equations were determined by Chandrasekhar \cite{Ch1} and Weinberg \cite{Wein} and the second post-Newtonian hydrodynamic equations by Chandrasekhar and Nutku \cite{ChNu}.

The derivation of hydrodynamic equations from a transfer equation for arbitrary macroscopic quantities which are associated with mean values of microscopic quantities is an old subject in the literature  of kinetic theory of gases which goes back to the work of Maxwell \cite{Max} in 1867 . In  1911 Enskog   \cite{Ens}  determined  from the Boltzmann equation a general transfer equation for an arbitrary function of the space-time and particle velocity  where the hydrodynamic equations can be obtained.

The main objective of this work is to develop a kinetic theory of gases within the framework of the second post-Newtonian approximation. The first post-Newtonian version of Boltzmann equation was proposed in the works \cite{Rez, Ped}. Here the second post-Newtonian approximation for the Boltzmann equation is determined from the evolution of the one-particle distribution function with respect to the proper time along the world line of the particle. 

In  equilibrium the relativistic gas  is characterized  by the Maxwell-J\"uttner distribution function (see e.g. \cite{CK}) here its expression in the second post-Newtonian approximation is obtained from the components of the metric tensor and of the hydrodynamic and particle four-velocities in the same order.    From the knowledge of the  Maxwell-J\"uttner distribution function  it is possible to determine the second post-Newtonian approximation of  the particle four-flow and energy-momentum tensor components through integration of their expressions which are  defined in terms of the one-particle distribution function. The resulting  expressions  correspond to those obtained from a phenomenological theory based on their decomposition in terms of the hydrodynamic four-velocity.  The components of the particle four-flow and energy-momentum tensor in the first post-Newtonian approximation  from the  Maxwell-J\"uttner distribution function were determined in \cite{KRW}.

From the  usual method in the kinetic theory of gases the hydrodynamic equations for the mass density, mass-energy density and momentum density are obtained from the integration of the Boltzmann equation when it is multiplied with the rest mass and the components of the particle momentum four-vector  -- which corresponds in the non-relativistic case to the energy and momentum of a particle --  respectively.  Due to the conservation laws of  the particle rest mass and momentum four-vector  these quantities  are identified as summational invariants which imply vanishing production terms related  to the collision operator of the Boltzmann equation. The Eulerian hydrodynamic equations derived from the post-Newtonian Boltzmann equation correspond to those obtained from the phenomenological theory which follow from the conservation equations  of the particle four-flow and energy-momentum tensor \cite{ChNu}. By neglecting the relativistic corrections the hydrodynamic equations for the mass  and mass-energy densities coincide and  correspond to the Newtonian continuity equation. However, their difference  leads to the hydrodynamic equation for the internal energy density, which is an expression in the first post-Newtonian approximation.  This result is compatible with  energy conservation law in the post-Newtonian theory \cite{Ch1,Ch2,ChNu}, since the first post-Newtonian expression for the energy conservation law follows only from the knowledge of the second post-Newtonian approximation. 

The paper  is outlined as follows: in Section \ref{s.2} we introduce the main results from the second post-Newtonian approximation theory that will be used in the paper. The determination of the second post-Newtonian Boltzmann equation, Maxwell-J\"uttner distribution function and Eulerian hydrodynamic equations are the subject of the Sections \ref{s.3}, \ref{s.4} and \ref{s.5}, respectively. The conclusions of the work are stated in the last section. 
The notations used here are: Greek indices take the values 0,1,2,3 and Latin indices the values 1,2,3. The semicolon denotes the covariant differentiation, the indices of Cartesian tensors will be written as subscripts, the summation convention over repeated  indices will be assumed and the partial differentiation will be denoted by $\partial/\partial x^i$.

\section{Preliminaries}
\lb{s.2}

In this work we shall analyse a relativistic perfect fluid characterized by the fields of particle four-flow $N^\mu$ and the energy-momentum tensor $T^{\mu\nu}$. These fields are usually decomposed in terms of the four-velocity $U^\mu$ (such that $U^\mu U_\mu=c^2$) as
\ben\lb{sp01}
N^\mu=nU^\mu,\qquad
 T^{\mu\nu}=(\epsilon+p)\frac{U^\mu U^\nu}{c^2}-pg^{\mu\nu}.
\een
Here $n$, $p$ and $\epsilon$  denote the particle number density,   the pressure and the energy density of the relativistic fluid, respectively. The energy density   has two parts $\epsilon=\rho c^2(1+\varepsilon/c^2)$ one associated with the mass density $\rho=mn$ -- where $m$ denotes the rest mass of a fluid particle  -- and another to the internal energy density $\rho\varepsilon$ of the gas.   In the work of Chandrasekhar \cite{Ch1} the internal energy density $\rho\varepsilon$ is represented by $\rho\Pi$ while in the book of Weinberg \cite{Wein} the energy density $\epsilon$ is simply represented  by $\rho$.

The post-Newtonian approximation is  a method for the solution of Einstein's field equations by considering successive approximations which are  expanded in powers of $1/c^2$. The starting point is the general expression for the line element $ds$ written in terms of the metric tensor $g_{\mu\nu}$, namely
$ds^2=c^2d\tau^2=g_{\mu\nu}dx^\mu dx^\nu=g_{00}(dx^0)^2+2g_{0i}dx^0dx^i+g_{ij}dx^idx^j,$
where $\tau$ is the proper time and $dx^0=cdt$.
From the knowledge  of the metric tensor components in a Minkowski space-time $g_{00}=1$, $g_{ij}=-\delta_{ij}$ and $g_{0i}=0$  the components of the metric tensor are split as
\ben\lb{sp02}
g_{00}=1+{\buildrel\!\!\!\! _{2} \over{g_{00}}}+{\buildrel\!\!\!\! _{4} \over{g_{00}}}+{\buildrel\!\!\!\! _{6} \over{g_{00}}}+\mathcal{O}(c^{-8}),
\qquad
g_{ij}=-\delta_{ij}+{\buildrel\!\!\!\! _{2} \over{g_{ij}}}+{\buildrel\!\!\!\! _{4} \over{g_{ij}}}+\mathcal{O}(c^{-6}),
\qquad
g_{0i}={\buildrel\!\!\!\! _{3} \over{g_{0i}}}+{\buildrel\!\!\!\! _{5} \over{g_{0i}}}+\mathcal{O}(c^{-7}),
\een
where ${\buildrel\!\!\!\! _{n} \over{g_{\mu\nu}}}$ denote the metric tensor components of   order $\mathcal{O}(c^{-n})$.
Furthermore, the components of the Christoffel symbol, the Ricci tensor and the energy-momentum tensor are also split in orders $\mathcal{O}(c^{-n})$. 

The solution of Einstein's field equations leads the following expressions for the metric tensor  components\footnote{The correspondence of the potentials given here with those in  \cite{ChNu} are: $\Pi_i\rightarrow P_i$, $\Psi_{ij}\rightarrow Q_{ij}$, $\Psi_{0i}\rightarrow Q_{0i}$ and $\Psi_{00}\rightarrow Q_{00}$.}  (see Chandrasekhar and Nutku \cite{ChNu})
\ben\lb{sp1a}
&&g_{00}=1-\frac{2U}{c^2}+\frac2{c^4}\left(U^2-2\Phi\right)+ \frac{\Psi_{00}}{c^6}+\mathcal{O}(c^{-8}),
\\\lb{sp1b}
&&g_{0i}=\frac{\Pi_i}{c^3}+\frac{\Psi_{0i}}{c^5}+\mathcal{O}(c^{-7}),
\\\lb{sp1c}
&&g_{ij}=-\left(1+\frac{2U}{c^2}\right)\delta_{ij}+\frac{\Psi_{ij}}{c^4}+\mathcal{O}(c^{-6}),
\een
while the corresponding components of the Christoffel symbols are given in the Appendix A.

The Newtonian   $U$ and the post-Newtonian $\Phi$, $\Pi_i$, $\Psi_{ij},$ $\Psi_{0i}$ and $\Psi_{00}$ gravitational potentials that follow from Einstein's field equations are given in terms of the Poisson equations (see \cite{ChNu})
\ben\lb{sp2a}
&&\nabla^2U=-4\pi G\rho,\qquad\nabla^2\Phi=-4\pi G\rho\left(V^2+U+\frac\varepsilon2+\frac{3p}{2\rho}\right),
\qquad
\nabla^2\Pi_i=-16\pi G\rho V_i+\frac{\partial^2U}{\partial t\partial x^i},
\\\no
&&\nabla^2\Psi_{ij}=16\pi G\rho\bigg(V_iV_j-V^2\delta_{ij}-2\frac{p}\rho\delta_{ij}\bigg)-2\bigg(\delta_{ij}\nabla^2+\frac{\partial^2}{\partial x^i\partial x^j}\bigg)(U^2+2\Phi)
+4\frac{\partial U}{\partial x^i}\frac{\partial U}{\partial x^j}-2\frac{\partial^2U}{\partial t^2}\delta_{ij}
\\\lb{sp2b}
&&\qquad-\frac{\partial}{\partial t}\bigg(\frac{\partial \Pi_i}{\partial x^j}+\frac{\partial \Pi_j}{\partial x^i}\bigg),
\\\lb{sp2c}
&&\nabla^2\Psi_{0i}=-16\pi G\rho\left[V_i\left(V^2+\varepsilon+\frac{p}\rho+4U\right)-\frac{\Pi_i}2\right]-10\frac{\partial U}{\partial t}\frac{\partial U}{\partial x^i}
-2\frac{\partial U}{\partial x^j}\frac{\partial \Pi_j}{\partial x^i}+2\Pi_j\frac{\partial^2 U}{\partial x^i\partial x^j},
\\\no
&&\nabla^2\Psi_{00}=16\pi G\rho\left[V^2\left(V^2+\varepsilon+\frac{p}\rho+4U\right)-U^2-2\Phi\right]
+2\frac{\partial U}{\partial x^i}\frac{\partial \Pi_i}{\partial t}-6\left(\frac{\partial U}{\partial t}\right)^2+12\frac{\partial U}{\partial x^i}\frac{\partial \Phi}{\partial x^i}
\\\lb{sp2d}
&&\qquad-12U\left(\frac{\partial U}{\partial x^i}\right)^2+\frac{\partial \Pi_j}{\partial x^i}\left(\frac{\partial \Pi_i}{\partial x^j}-\frac{\partial \Pi_j}{\partial x^i}\right)+2\Psi_{ij}\frac{\partial^2 U}{\partial x^i\partial x^j}.
\een
In the above equations  $V_i$ is the fluid hydrodynamic velocity and in the Poisson equations for $\Psi_{ij},$ $\Psi_{0i}$ and $\Psi_{00}$ the gauge proposed in the work \cite{ChNu} was taken into account.

The components of the hydrodynamic four-velocity
\ben\lb{sp3a}
(U^\mu)=\left(\frac{dx^0}{d\tau}, \frac{dx^i}{d\tau}\right)=\left(c\frac{dt}{d\tau},V_i\frac{dt}{d\tau}\right),
\een
are determined from the  the expression
\ben\lb{sp3b}
\left(\frac{d\tau}{dt}\right)^2=g_{00}+\frac2cg_{0i}V^i+\frac1{c^2}g_{ij}V^iV^j=1-\frac1{c^2}\left(2U+V^2\right)
+\frac2{c^4}\left(U^2-UV^2-2\Phi+\Pi_iV^i\right)+\mathcal{O}(c^{-5}),
\een
which results from the line element $ds$ together  with the components of the metric tensor (\ref{sp1a}) -- (\ref{sp1c}).
Indeed, by using the approximation $1/{\sqrt{1+x}}\approx 1-{x}/2+{3x^2}/8$  the components of the hydrodynamic four-velocity read
\ben\lb{sp4}
U^0=c\left[1+\frac1{c^2}\left(\frac{V^2}2+U\right)
+\frac1{c^4}\left(\frac{3V^4}8+\frac{5U V^2}2+\frac{U^2}2+2\Phi-\Pi_iV_i\right)\right],
\qquad
U^i=\frac{V_iU^0}c.
\een

Once the components of the four-velocity (\ref{sp4}) are known up to order $1/c^4$ we can obtain the components of the particle four-flow and energy-momentum tensor (\ref{sp01}) in the corresponding order. They read
\ben\lb{sp5a}
&&N^0= nc\left[1+\frac1{c^2}\left(\frac{V^2}2+U\right)
+\frac1{c^4}\left(\frac{3V^4}8+\frac{5U V^2}2+\frac{U^2}2+2\Phi-\Pi_iV_i\right)\right],\qquad N^i=\frac{N^0V_i}c.
\\\lb{sp5b}
&&T^{00}= \rho c^2\left[1+\left(V^2+\varepsilon+2U\right)+\frac\rho{c^4}V^2\left(V^2+\varepsilon+\frac{p}\rho+6U\right)+2U\varepsilon-2\Pi_iV_i+2U^2+4\Phi\right],\qquad
\\\lb{sp5c}
&&T^{i0}=\rho cV_i\left\{1+\frac1{c^2}\left(V^2+2U+\varepsilon+\frac{p}\rho\right)+\frac1{c^4}\left[V^4+6V^2U+2U^2+\left(\varepsilon+\frac{p}\rho\right)(V^2+2U)4\Phi-2\Pi_jV_j\right]\right\}-\frac{p\Pi_i}{c^3},\qquad
\\\no
&&T^{ij}=\rho\left(V_iV_j+\frac{p}\rho\delta_{ij}\right)+\frac\rho{c^2}\bigg[\bigg(V^2
+2U+\varepsilon+\frac{p}\rho\bigg)V_iV_j -2\frac{p}\rho U\delta_{ij}\bigg]+\frac\rho{c^4}\bigg\{\bigg[V^4+6UV^2+2U^2
\\\lb{sp5d}
&&\qquad+\bigg(\varepsilon+\frac{p}\rho\bigg)(V^2+2U)-2\Pi_iV_i+4\Phi\bigg]V_iV_j
+\frac{p}\rho(4U^2\delta_{ij}+\Psi_{ij})\bigg\}.
\een

\section{Post-Newtonian Boltzmann Equation}
\lb{s.3}

In the relativistic kinetic theory a particle of a  gas  with rest mass $m$ is characterized by the space-time coordinates $(x^\mu)=(ct,\bx)$ and momentum four-vector $(p^\mu)=(p^0,\bp)$. From the constraint that the length of the momentum four-vector is constant, i.e. $g_{\mu\nu}p^\mu p^\nu=m^2c^2$, the time component is given in terms of its spatial components by
\ben\lb{k1}
p^0=\frac{\sqrt{g_{00}m^2c^2+(g_{0i}g_{0j}-g_{00}g_{ij})p^ip^j}-g_{0i}p^i}{g_{00}}.
\een

The state of the gas in the phase space represented by the spatial  coordinates $\bx$ and momentum  $\bp$ is characterized by the  one-particle distribution function $f(\bx,\bp,t)$ such that
$dN=f(\bx,\bp,t)d^3Xd^3P,$
gives the number of particle world lines that crosses the hypersurface element represented by the three-dimensional space $d^3X$ on the surface $x^0=$ constant and with the spatial momentum four-vector contained in the cell $d^3P$ of the mass-shell. The one-particle distribution function is a scalar invariant and the  invariant volume elements $d^3X$ and $d^3P$ are (see e.g. \cite{CK})
\ben\lb{k2}
d^3X=p^0\sqrt{-g}d^3x=p^{\prime0}\sqrt{-g'}d^3x',\qquad
d^3P=\sqrt{-g}\frac{d^3p}{p_0}=\sqrt{-g'}\frac{d^3p'}{p'_0},
\een
where $g$ is the determinant of the metric tensor.

The Boltzmann equation is a non-linear integro-differential equation for the space-time evolution of the one-particle distribution function $f(\bx,\bp,t)$ in the phase space spanned by the  spatial coordinates $\bx$ and momentum $\bp$ of the particles. Its  expression for collisionless systems in the first post-Newtonian approximation was  derived in the paper \cite{Rez,Ped}. Here we follow a different method for the derivation of the post-Newtonian Boltzmann equation, which was used for the determination of the Boltzmann equation in gravitational fields (see e.g. \cite{CK}).

We start by writing the one-particle distribution function $f(\bx,\bp,t)$ as $f(x^\mu(\tau),v_i(\tau))$ where $\bv=\bp/m$ represents the particle velocity  and  $\tau$  the proper time along the world line of the particle. The variation of the one-particle distribution function with respect to the proper time is
\ben\lb{k3}
\frac{df(x^\mu(\tau),v_i(\tau))}{d\tau}=\frac{\partial f}{\partial x^\mu}\frac{dx^\mu}{d\tau}+\frac{\partial f}{\partial v_i}\frac{dv_i}{d\tau}=u^\mu\frac{\partial f}{\partial x^\mu}+\frac{\partial f}{\partial v_i}\frac{dv_i}{dt}\frac{dt}{d\tau},
\een
where $(u^\mu)=(u^0,u^0v_i/c)$ is the four-velocity of the gas particles. The corresponding expressions for $u^0$ and $u^i$ are obtained from the substitution  of the hydrodynamic velocities $U^\mu$ and $\bf V$ by the particle velocities $u^\mu$ and $\bv$ in (\ref{sp4}), namely
\ben\lb{k3a}
u^0=c\left[1+\frac1{c^2}\left(\frac{v^2}2+U\right)
+\frac1{c^4}\left(\frac{3v^4}8+\frac{5U v^2}2+\frac{U^2}2+2\Phi-\Pi_iv_i\right)\right],
\qquad
u^i=\frac{v_iu^0}c.
\een

The first term of the second equality in (\ref{k3}) computed to the $1/c^4$ order is
\ben\lb{k4}
u^\mu\frac{\partial f}{\partial x^\mu}=u^0\frac{\partial f}{\partial x^0}+u^i\frac{\partial f}{\partial x^i}=\left[1+\frac1{c^2}\left(\frac{v^2}2+U\right)
+\frac1{c^4}\left(\frac{3v^4}8+\frac{5U v^2}2+\frac{U^2}2+2\Phi-\Pi_iv_i\right)\right]\left(\frac{\partial f}{\partial t}+v_i\frac{\partial f}{\partial x^i}\right).
\een

For the second term of the last equality in (\ref{k3}) 
we calculate first  the equation of motion of the gas particles
\ben\lb{k5}
\frac{d^2x^\mu}{d\tau^2}+{\Gamma^\mu}_{\nu\lambda}\frac{dx^\nu}{d\tau}\frac{dx^\lambda}{d\tau}=0,
\een
and compute the acceleration which follows from  this equation  (see Weinberg \cite{Wein})
\ben\lb{k6}
&&\frac{d^2 x^i}{d(x^0)^2}=\left(\frac{dx^0}{d\tau}\right)^{-1}\frac{d}{d\tau}\left[\left(\frac{dx^0}{d\tau}\right)^{-1}\frac{dx^i}{d\tau}\right]
=\left(\frac{dx^0}{d\tau}\right)^{-2}\bigg[\frac{d^2x^i}{d\tau^2}-\left(\frac{dx^0}{d\tau}\right)^{-1}\frac{d^2x^0}{d\tau^2}\frac{dx^i}{d\tau}\bigg].
\een
This  equation can be rewritten by using (\ref{k5}) as
\ben\no
&&\frac{d^2 x^i}{d(x^0)^2}=\left(\frac{dx^0}{d\tau}\right)^{-2}\frac{dx^\mu}{d\tau}
\frac{dx^\nu}{d\tau}\bigg[{\Gamma^0}_{\mu\nu}\left(\frac{dx^0}{d\tau}\right)^{-1}\frac{dx^i}{d\tau}
-{\Gamma^i}_{\mu\nu}\bigg]
=-{\Gamma^i}_{00}+{\Gamma^i}_{jk}\frac{dx^j}{dx^0}\frac{dx^k}{dx^0}-2{\Gamma^i}_{0j}\frac{dx^j}{dx^0}
\\\lb{k7}
&&\qquad+\frac{dx^i}{dx^0}\left[{\Gamma^0}_{00}+2{\Gamma^0}_{0j}\frac{dx^j}{dx^0}
+{\Gamma^0}_{jk}\frac{dx^j}{dx^0}\frac{dx^k}{dx^0}\right].\,
\een
If we use in (\ref{k7}) the expressions for the Christoffel symbol components in the Appendix A  its post-Newtonian approximation up to order $1/c^{6}$  becomes
\ben\no
&&\frac{d^2 x^i}{dt^2}=c^2\bigg\{-{\buildrel\!\!\!\! _{2} \over{\Gamma^i}_{00}}-{\buildrel\!\!\!\! _{4} \over{\Gamma^i}_{00}}-{\buildrel\!\!\!\! _{6} \over{\Gamma^i}_{00}}-2\frac{v_j}{c}\bigg({\buildrel\!\!\!\! _{3} \over{\Gamma^i}_{0j}}+{\buildrel\!\!\!\! _{5} \over{\Gamma^i}_{0j}}\bigg)+\frac{v_i}c\bigg[{\buildrel\!\!\!\! _{3} \over{\Gamma^0}_{00}}+{\buildrel\!\!\!\! _{5} \over{\Gamma^0}_{00}}+2\frac{v_j}c\bigg({\buildrel\!\!\!\! _{2} \over{\Gamma^0}_{0j}}+{\buildrel\!\!\!\! _{4} \over{\Gamma^0}_{0j}}\bigg)+{\frac{v_j v_k}{c^2}{\buildrel\!\!\!\! _{3} \over{\Gamma^0}_{jk}}}\bigg]
\\\no
&&-\frac{v_j v_k}{c^2}\bigg({\buildrel\!\!\!\! _{2} \over{\Gamma^i}_{jk}}+{\buildrel\!\!\!\! _{4} \over{\Gamma^i}_{jk}}\bigg)\bigg\}
=\frac{\partial U}{\partial x^i}-\frac{v_i}{c^2}\bigg[\frac{\partial U}{\partial t}+2v_j\frac{\partial U}{\partial x^j}\bigg]-\frac1{c^2}\left(1-\frac{2U}{c^2}\right)\bigg[2\frac{\partial\left(U^2-\Phi\right)}
{\partial x^i}+\bigg(2v_iv_j\frac{\partial U}{\partial x^j}-v^2\frac{\partial U}{\partial x^i}\bigg)
\\\no
&&-\frac{\partial\Pi_i}{\partial t}-v_j\bigg(\frac{\partial\Pi_i}{\partial x^j}-\frac{\partial\Pi_j}{\partial x^i}-2\delta_{ij}\frac{\partial U}{\partial t}\bigg)\bigg]
+\frac{v_i}{c^4}\bigg[v_jv_k\frac{\partial \Pi_j}{\partial x^k}+v^2\frac{\partial U}{\partial t}\bigg]+\frac{2v_j}{c^4}\bigg[\Pi_i\frac{\partial U}{\partial x^j}+\frac12\frac{\partial \Psi_{i0}}{\partial x^j}
-\frac12\frac{\partial \Psi_{j0}}{\partial x^i}+\frac12\frac{\partial \Psi_{ij}}{\partial t}\bigg]
\\\lb{k8}
&&+\frac{v_i}{c^4}\bigg[\Pi_j\frac{\partial U}{\partial x^j}-2\frac{\partial \Phi}{\partial t}-4v_j\frac{\partial \Phi}{\partial x^j}\bigg]+\frac{v_jv_k}{2c^4}\bigg[2\frac{\partial \Psi_{ij}}{\partial x^k}-\frac{\partial \Psi_{jk}}{\partial x^i}\bigg]+\frac1{c^4}\bigg[\Pi_i\frac{\partial U}{\partial t}+\frac{\partial \Psi_{i0}}{\partial t}-\frac12\frac{\partial \Psi_{00}}{\partial x^i}+\Psi_{ij}\frac{\partial U}{\partial x^j}\bigg].
\een

The Boltzmann equation in the second post-Newtonian approximation follows from (\ref{k3}) by taking into account (\ref{k4}) and (\ref{k8}) and reads
\ben\lb{k9}
&&\bigg[\frac{\partial f}{\partial t}+v_i\frac{\partial f}{\partial x^i}+\frac{\partial f}{\partial v_i}\frac{\partial U}{\partial x^i}\bigg]\bigg[1+\frac1{c^2}\left(\frac{v^2}2+U\right)+\frac1{c^4}\bigg(\frac{3v^4}8
+\frac{5v^2U}2+\frac{U^2}2+2\Phi-\Pi_jv_j\bigg)\bigg]
\\\no
&&+\frac1{c^2}\frac{\partial f}{\partial v_i}\bigg\{\left[1+\frac1{c^2}\left(\frac{v^2}2
-U\right)\right]\bigg[v_j\bigg(\frac{\partial\Pi_i}{\partial x^j}
-\frac{\partial\Pi_j}{\partial x^i}\bigg)-2v_i\frac{\partial U}{\partial t}+\frac{\partial\Pi_i}{\partial t}-2\frac{\partial\left(U^2-\Phi\right)}{\partial x^i}-2v_iv_j\frac{\partial U}{\partial x^j}+v^2\frac{\partial U}{\partial x^i}\bigg]
\\\no
&&-v_i\bigg[1+\frac1{c^2}\left(\frac{v^2}2+U\right)\bigg]
\bigg[\frac{\partial U}{\partial t}+2v_j\frac{\partial U}{\partial x^j}\bigg]+\frac{v_jv_k}{2c^2}\bigg[2\frac{\partial \Psi_{ij}}{\partial x^k}-\frac{\partial \Psi_{jk}}{\partial x^i}\bigg]+\frac1{c^2}\bigg[\Pi_i\frac{\partial U}{\partial t}
+\frac{\partial \Psi_{i0}}{\partial t}-\frac12\frac{\partial \Psi_{00}}{\partial x^i}+\Psi_{ij}\frac{\partial U}{\partial x^j}\bigg]
\\\no
&&+\frac{v_i}{c^2}\bigg[v_jv_k\frac{\partial \Pi_j}{\partial x^k}
+v^2\frac{\partial U}{\partial t}\bigg]
+\frac{2v_j}{c^2}\bigg[\Pi_i\frac{\partial U}{\partial x^j}+\frac12\frac{\partial \Psi_{i0}}{\partial x^j}-\frac12\frac{\partial \Psi_{j0}}{\partial x^i}+\frac12\frac{\partial \Psi_{ij}}{\partial t}\bigg]
+\frac{v_i}{c^2}\bigg[\Pi_j\frac{\partial U}{\partial x^j}-2\frac{\partial \Phi}{\partial t}
-4v_j\frac{\partial \Phi}{\partial x^j}\bigg]\bigg\}=\mathcal{Q}(f,f).
\een
Here we have introduced  the so-called collision operator of the Boltzmann equation $\mathcal{Q}(f,f)$, which refers to the binary collision of the particles and is given in terms of an integral of the product of two particle distribution functions at collision.
In the first post-Newtonian approximation the Boltzmann equation reduces to
\ben\no
&&\bigg[\frac{\partial f}{\partial t}+v_i\frac{\partial f}{\partial x^i}+\frac{\partial f}{\partial v_i}\frac{\partial U}{\partial x^i}\bigg]\bigg[1+\frac1{c^2}\left(\frac{v^2}2+U\right)\bigg]+\frac1{c^2}\frac{\partial f}{\partial v_i}\bigg\{v_j\bigg(\frac{\partial\Pi_i}{\partial x^j}
-\frac{\partial\Pi_j}{\partial x^i}\bigg)-2v_i\frac{\partial U}{\partial t}
\\\lb{k9a}
&&\qquad+\frac{\partial\Pi_i}{\partial t}-2\frac{\partial\left(U^2-\Phi\right)}{\partial x^i}-2v_iv_j\frac{\partial U}{\partial x^j}+v^2\frac{\partial U}{\partial x^i}-v_i
\bigg[\frac{\partial U}{\partial t}+2v_j\frac{\partial U}{\partial x^j}\bigg]
\bigg\}=\mathcal{Q}(f,f),
\een
which is the expression obtained in \cite{Ped} if we identify there $\phi\rightarrow -U$, $\xi_i\rightarrow-\Pi_i$ and $\psi\rightarrow-2\Phi.$

\section{Post-Newtonian Maxwell-J\"uttner Distribution Function}\lb{s.4}

At equilibrium the collision term of the Boltzmann vanishes, since it refers to the difference of the number of particles that enter and leave the volume element  in the phase space. An important consequence is that  one can determine the equilibrium distribution function from the condition that the collision term vanishes at equilibrium (for more details one is referred to e.g. 
see \cite{CK}). In the relativistic kinetic theory the equilibrium distribution function is the so-called Maxwell-J\"uttner distribution function which reads
\ben\lb{mj1}
f(\bx,\bp,t)=\frac{n}{4\pi m^2ckTK_2(\zeta)}\exp\left(-\frac{p^\mu U_\mu}{kT}\right).
\een
Here $k$ is the Boltzmann constant, $T$ the absolute temperature and $K_2(\zeta)$ the modified Bessel function of the second kind which depends on the relativistic parameter $\zeta$. The relativistic parameter $\zeta=mc^2/kT$ represents the ratio of the rest energy of the gas particles $mc^2$ and the thermal energy of the gas $kT$ and in the non-relativistic limiting case $\zeta\gg1$ while in the ultra-relativistic limiting case $\zeta\ll1$.

For the determination of the Maxwell-J\"uttner distribution function in the second post-Newtonian approximation we begin with the determination of its exponential term. We introduce the so-called peculiar velocity $\mathcal{V}_i=v_i-V_i$ --  which is the difference of the particle velocity and the gas velocity, i.e. it refers to the particle velocity  in the gas frame -- and get the following relationship by using the components of the metric tensor $g_{\mu\nu}$, fluid $U^\mu$ and particle $u^\mu=p^\mu/m$ four-velocities
\ben\no
&&\frac{g_{\mu\nu}p^\mu U^\nu}{kT}=\frac{m}{kT}\bigg\{c^2+\frac{\mathcal{V}^2}{2}+\frac{1}{c^2} \bigg[2 U \mathcal{V}^2+\frac{V^2 \mathcal{V}^2}{2}+\frac{(V_i\mathcal{V}_i)^2}{2}+(V_i\mathcal{V}_i) \mathcal{V}^2+\frac{3 \mathcal{V}^4}{8}\bigg]
+\frac{1}{c^4} \bigg[3 U^2 \mathcal{V}^2+4 U V^2 \mathcal{V}^2
\\\no
&&\quad+4 U (V_i\mathcal{V}_i)^2+8 U (V_i\mathcal{V}_i) \mathcal{V}^2+3 U \mathcal{V}^4+\frac{V^4 \mathcal{V}^2}{2}+V^2 (V_i\mathcal{V}_i)^2+2 V^2 (V_i\mathcal{V}_i) \mathcal{V}^2
+\frac{3 V^2 \mathcal{V}^4}{4}+(V_i\mathcal{V}_i)^3
\\\lb{mj2}
&&\quad+\frac{9 (V_i\mathcal{V}_i)^2 \mathcal{V}^2}{4}+\frac{3 (V_i\mathcal{V}_i) \mathcal{V}^4}{2}+\frac{5 \mathcal{V}^6}{16}-\Pi_i\mathcal{V}_i (V_i\mathcal{V}_i)-\Pi_i V_i \mathcal{V}^2-\Pi_i \mathcal{V}_i \mathcal{V}^2+2 \Phi  \mathcal{V}^2-\frac{\mathcal{V}_i \mathcal{V}_j \Psi_{ij}}{2}\bigg]\bigg\}.
\een
Next up to the $1/c^4$ order the modified Bessel function of second kind reads \cite{AbSt}
\ben\lb{mj3}
\frac1{K_2(\zeta)}=\sqrt{\frac{2mc^2}{\pi kT}}\,e^{\frac{mc^2}{kT}}
\left(1-\frac{15kT}{8mc^2}+\frac{345(kT)^2}{128m^2c^4} \right).
\een

The  Maxwell-J\"uttner distribution function in the second post-Newtonian approximation is obtained from (\ref{mj1}) -- (\ref{mj3}) by considering the approximation $e^{-x}\approx1-x+x^2/2$ for the  $1/c^2$ -- terms in the exponential, yielding
\ben\nonumber
&&f=\frac{n}{(2\pi mkT)^{\frac32}}e^{-\frac{m\mathcal{V}^2}{2kT}}
\Bigg\{ 1-\frac{1}{c^2}\bigg[\frac{15 kT}{8 m}+\frac{m (V_i\mathcal{V}_i)^2}{2 kT}+\frac{2 m U \mathcal{V}^2}{kT}+\frac{3 m \mathcal{V}^4}{8 kT}+\frac{m V^2 \mathcal{V}^2}{2 kT}+\frac{m (V_i\mathcal{V}_i) \mathcal{V}^2}{kT}\bigg]
\\\no
&&+\frac{1}{c^4}\bigg[\frac{2 m^2 U^2 \mathcal{V}^4}{(kT)^2}+\frac{m^2 U V^2 \mathcal{V}^4}{(kT)^2}+\frac{3 m^2 U \mathcal{V}^6}{4 (kT)^2}
+\frac{m^2 U (V_i\mathcal{V}_i)^2 \mathcal{V}^2}{(kT)^2}+\frac{2 m^2 U (V_i\mathcal{V}_i) \mathcal{V}^4}{(kT)^2}+\frac{3 m^2 V^2 \mathcal{V}^6}{16 (kT)^2}+\frac{m^2 V^4 \mathcal{V}^4}{8 (kT)^2}
\\\no
&&+\frac{m^2 (V_i\mathcal{V}_i)^4}{8 (kT)^2}+\frac{m^2 V^2 (V_i\mathcal{V}_i)^2 \mathcal{V}^2}{4 (kT)^2}+\frac{m^2 V^2 (V_i\mathcal{V}_i) \mathcal{V}^4}{2 (kT)^2}+\frac{m^2 (V_i\mathcal{V}_i)^3 \mathcal{V}^2}{2 (kT)^2}+\frac{11 m^2 (V_i\mathcal{V}_i)^2 \mathcal{V}^4}{16 (kT)^2}+\frac{3 m^2 (V_i\mathcal{V}_i) \mathcal{V}^6}{8 (kT)^2}
\\\no
&&+\frac{9 m^2 \mathcal{V}^8}{128 (kT)^2}+\frac{345 (kT)^2}{128 m^2}
-\frac{3 m U^2 \mathcal{V}^2}{kT}-\frac{4 m U V^2 \mathcal{V}^2}{kT}-\frac{4 m U (V_i\mathcal{V}_i)^2}{kT}-\frac{8 m U (V_i\mathcal{V}_i) \mathcal{V}^2}{kT}-\frac{3 m U \mathcal{V}^4}{kT}+\frac{m \Pi_i\mathcal{V}_i (V_i\mathcal{V}_i)}{kT}
\\\no
&&-\frac{m V^4 \mathcal{V}^2}{2 kT}-\frac{m V^2 (V_i\mathcal{V}_i)^2}{kT}
-\frac{2 m V^2 (V_i\mathcal{V}_i) \mathcal{V}^2}{kT}-\frac{3 m V^2 \mathcal{V}^4}{4 kT}-\frac{m (V_i\mathcal{V}_i)^3}{kT}-\frac{9 m (V_i\mathcal{V}_i)^2 \mathcal{V}^2}{4 kT}-\frac{3 m (V_i\mathcal{V}_i) \mathcal{V}^4}{2 kT}-\frac{5 m \mathcal{V}^6}{16 kT}
\\\lb{mj4}
&&+\frac{m \Pi_i V_i \mathcal{V}^2}{kT}
+\frac{m \Pi_i \mathcal{V}_i \mathcal{V}^2}{kT}-\frac{2 m \Phi  \mathcal{V}^2}{kT}+\frac{m \mathcal{V}_i \mathcal{V}_j \Psi_{ij}}{2 kT}+\frac{15 U \mathcal{V}^2}{4}+\frac{15 V^2 \mathcal{V}^2}{16}
+\frac{15 (V_i\mathcal{V}_i)^2}{16}+\frac{15 (V_i\mathcal{V}_i) \mathcal{V}^2}{8}+\frac{45 \mathcal{V}^4}{64}\bigg]\bigg\}.
\een
In the first post-Newtonian approximation the Maxwell-J\"uttner distribution function becomes \cite{KRW}
\ben
f=\frac{n}{(2\pi mkT)^{\frac32}}e^{-\frac{m\mathcal{V}^2}{2kT}}
\Bigg\{ 1-\frac{1}{c^2}\bigg[\frac{15 kT}{8 m}+\frac{m (V_i\mathcal{V}_i)^2}{2 kT}+\frac{2 m U \mathcal{V}^2}{kT}+\frac{3 m \mathcal{V}^4}{8 kT}+\frac{m V^2 \mathcal{V}^2}{2 kT}+\frac{m (V_i\mathcal{V}_i) \mathcal{V}^2}{kT}\bigg]\bigg\}.
\een

In the kinetic theory of relativistic gases  the particle four-flow and the energy-momentum tensor are given in terms of the one-particle distribution function $f(\bx,\bv,t)$ (see e.g. \cite{CK})  by
\ben\lb{mj5}
N^\mu=m^4c\int u^\mu  f\frac{\sqrt{-g}\,d^3 u}{u_0},
\qquad T^{\mu\nu}=m^4c\int u^\mu u^\nu f\frac{\sqrt{-g}\,d^3 u}{u_0}.
\een

The transformation of the differential elements $d^3u=du^1du^2du^3=\vert J\vert dv_1dv_2dv_3=\vert J\vert d^3v$ is given by the Jacobian matrix which follows from (\ref{k3a}) by computing of the derivatives $J=\partial(u^1,u^2,u^3)/\partial(v_1,v_2,v_3)$. The final expression for the transformation of the differential elements in the second post-Newtonian approximation reads
\ben\lb{mj6}
 d^3u=\bigg[1+\frac1{c^2}\bigg(\frac{5v^2}{2}+3U\bigg)+\frac1{c^4}\bigg(\frac{35v^4}8
 +\frac{9U^2}2+6\Phi
+\frac{35Uv^2}2-4\Pi_iv_i\bigg)
\bigg]d^3v.
\een
Moreover, from the expressions for the components of the metric tensor   we have that
\ben\lb{mj7}
u_0=(g_{00}u^0+g_{0i}u^i)
= u^0\left[1-2\frac{U}{c^2}+\frac1{c^4}\left(2U^2-4\Phi+\Pi_iv_i\right)\right],
\qquad
\sqrt{-g}=1+\frac{2U}{c^2}-\frac1{c^4}\left(U^2+2\Phi+\frac{\Psi_{kk}}2\right).
\een
Hence the integration element up to the  $1/c^4$ order which follows from (\ref{mj6}) and (\ref{mj7}) reads
\ben\no
&&\frac{\sqrt{-g}\, d^3 u}{u_0}=
\bigg[1+\frac1{c^2}\bigg(\frac{5v^2}{2}
+7U\bigg)+\frac1{c^4}\bigg(\frac{35v^4}8+\frac{55Uv^2}2
 +\frac{43U^2}2+8\Phi-\frac{\Psi_{kk}}2-5\Pi_iv_i\bigg)
\bigg]\frac{d^3v}{u^0}
\\\lb{mj8}
&&\qquad=\Bigg\{1+\frac1{c^2}\Bigg[2v^2+6U\Bigg]
+\frac1{c^4}\Bigg[3v^4+20Uv^2+15U^2+6\Phi-4\Pi_iv_i
-\frac{\Psi_{kk}}2\Bigg]\Bigg\}\frac{d^3v}c.
\een

From the knowledge of the  Maxwell-J\"uttner distribution function (\ref{mj4}) and of the integration element (\ref{mj8}) it is possible to determine the components of the particle four-flow and energy-momentum tensor in the second post-Newtonian approximation. 
For the integration it is necessary to introduce the peculiar velocity $\mathcal{V}_i=v_i-V_i$ in (\ref{mj8}), spherical coordinates and  express the integral element as $d^3\mathcal{V}=\mathcal{V}^2\sin\theta d\theta d\varphi$, where $0\leq\theta\leq\pi$ and $0\leq\varphi\leq2\pi$. 

The insertion of the Maxwell-J\"uttner distribution function (\ref{mj4}) and the integration element (\ref{mj8}) into the definition of the particle four-flow (\ref{mj5})$_1$ with subsequent  integration of the resulting equation leads to (\ref{sp5a}). Note that for the integration one makes use of the table of integrals given in the Appendix B.
The components of the energy-momentum tensor (\ref{mj5})$_2$ follow in the same manner and read
\ben\no
&&T^{00}=\rho c^2\bigg[1+\frac1{c^2}\bigg(V^2+2U+\frac{3kT}{2m}\bigg)+\frac1{c^4}\bigg(V^4+6UV^2+2U^2
+\frac{5kTV^2}{2m}+\frac{15(kT)^2}{8m^2}+\frac{3kTU}m-2\Pi_iV_i+4\Phi\bigg)\bigg],
\\\lb{mj9a}
\\\no
&&T^{0i}=\rho cV_i\bigg[1+\frac1{c^2}\bigg(V^2+2U+\frac{5kT}{2m}\bigg)+\frac1{c^4}\bigg(V^4+6UV^2+2U^2
+\frac{5kTV^2}{2m}+\frac{15(kT)^2}{8m^2}+\frac{5kTU}{m}
\\\lb{mj9c}
&&\qquad-2\Pi_jV_j+4\Phi\bigg)\bigg]-\frac{kT\Pi_i}{mc^4},
\\\no
&&T^{ij}=\rho\left(V_iV_j+\frac{kT}m\delta_{ij}\right)+\frac\rho{c^2}\bigg[\bigg(V^2+2U
+\frac{3kT_0}{2m}\bigg)V_iV_j -\frac{2kTU}m\delta_{ij}\bigg]
+\frac\rho{c^4}\bigg[\bigg(V^4+6UV^2+2U^2
\\\lb{mj9d}
&&\qquad+\frac{5kTV^2}{2m}+\frac{15(kT)^2}{8m^2}+\frac{5kTU}{m}-2\Pi_kV_k+4\Phi\bigg)V_iV_j+\frac{4kTU^2}{m}\delta_{ij}
+\frac{kT\Psi_{ij}}{m}\bigg].
\een
If we make use of the thermal equation of state and the expression of the specific internal energy, namely 
\ben\lb{mj9b}
 p=\frac{\rho kT}m,\qquad \varepsilon=\frac{3kT}{2m}\left(1+\frac{5kT}{4mc^2}\right),
\een
the above expressions for the  components of the energy-momentum tensor match the ones given by (\ref{sp5b}) -- (\ref{sp5d}).

\section{Post-Newtonian Eulerian Hydrodynamic Equations}\lb{s.5}

The   Eulerian hydrodynamic equations in the first post-Newtonian approximation were determined by  Chandrasekhar \cite{Ch1} and Weinberg \cite{Wein} from a  macroscopic description based on the equations of conservation of the particle four-flow $N^\mu$ and energy-momentum tensor $T^{\mu\nu}$, namely
\ben\lb{ba1}
{N^\mu}_{;\mu}=\frac{\partial N^\mu}{\partial x^\mu}+{\Gamma^\mu}_{\mu\lambda}N^\lambda=0,
\qquad
{T^{\mu\nu}}_{;\nu}=\frac{\partial T^{\mu\nu}}{\partial x^\nu}+{\Gamma^\mu}_{\nu\lambda}T^{\lambda\nu}+{\Gamma^\nu}_{\nu\lambda}T^{\mu\lambda}=0,
\een
while the  second post-Newtonian Eulerian hydrodynamic equations were obtained by Chandrasekhar and Nutku \cite{ChNu}. Here we shall obtain the second post-Newtonian Eulerian hydrodynamic equations from the Boltzmann equation (\ref{k9}). 

\subsection{Mass density hydrodynamic equation}

 We begin with the determination of the mass density hydrodynamic equation and for that end we multiply the  Boltzmann equation (\ref{k9}) by $m^4\sqrt{-g}d^3u/u_0$, use the Maxwell-J\"uttner distribution function (\ref{mj4}), the integration element (\ref{mj8}) and integrate the resulting equation, yielding
\ben\no
&&\frac{\partial}{\partial t}\bigg\{\rho\bigg[1+\frac1{c^2}\bigg(\frac{V^2}2+U\bigg)+\frac1{c^4}\bigg(\frac{3V^4}8+\frac{5V^2U}2
+\frac{U^2}2+2\Phi-\Pi_jV_j\bigg)\bigg]\bigg\}
\\\no
&&+\frac{\partial}{\partial x^i}\bigg\{\rho V_i\bigg[1+\frac1{c^2}\bigg(\frac{V^2}2+U\bigg)+\frac1{c^4}\bigg(\frac{3V^4}8+\frac{5V^2U}2
+\frac{U^2}2+2\Phi-\Pi_jV_j\bigg)\bigg]\bigg\}
\\\lb{ba2a}
&&+\underline{2\frac\rho{c^2}\bigg(\frac{\partial U }{\partial t}+V_i\frac{\partial U}{\partial x^i}\bigg)}+\underline{\frac{\rho}{c^4}\bigg(\frac{\partial U }{\partial t}+V_i\frac{\partial U}{\partial x^i}\bigg)\big(V^2-4U\big)}
-\underline{\frac\rho{c^4}\bigg[2\bigg(\frac{\partial \Phi }{\partial t}+V_i\frac{\partial \Phi}{\partial x^i}\bigg)+\frac12\bigg(\frac{\partial \Psi_{kk} }{\partial t}+V_i\frac{\partial \Psi_{kk}}{\partial x^i}\bigg)\bigg]}=0.
\een
The Newtonian continuity equation follows from the above equation by neglecting all terms in $1/c^2$ and $1/c^4$
\ben\lb{ba2b}
\frac{\partial\rho}{\partial t}+\frac{\partial\rho V_i}{\partial x^i}=0.
\een
The first post-Newtonian approximation for the continuity follows from (\ref{ba2a}) neglecting all $1/c^4$ terms
\ben\lb{ba2c}
\frac{\partial}{\partial t}\bigg\{\rho\bigg[1+\frac1{c^2}\bigg(\frac{V^2}2+U\bigg)\bigg]\bigg\}+\frac{\partial}{\partial x^i}\bigg\{\rho V_i\bigg[1+\frac1{c^2}\bigg(\frac{V^2}2+U\bigg)\bigg]\bigg\}+2\frac\rho{c^2}\bigg(\frac{\partial U }{\partial t}+V_i\frac{\partial U}{\partial x^i}\bigg)=0.
\een
The last term above can be rewritten as
\ben\lb{ba2d}
2\frac\rho{c^2}\bigg(\frac{\partial U }{\partial t}+V_i\frac{\partial U}{\partial x^i}\bigg)=\frac2{c^2}\bigg(\frac{\partial \rho U }{\partial t}+\frac{\partial \rho UV_i}{\partial x^i}\bigg)-\frac{2U}{c^2}\underline{\bigg(\frac{\partial \rho }{\partial t}+\frac{\partial \rho V_i}{\partial x^i}\bigg)},
\een
where for the underlined term vanishes thanks to the Newtonian continuity equation (\ref{ba2c}). Hence it follows the final form of the continuity equation in the first post-Newtonian approximation
\ben\lb{ba2e}
\frac{\partial\rho_*}{\partial t}+\frac{\partial\rho_* V_i}{\partial x^i}=0,\qquad \hbox{where}\qquad \rho_*=\rho\bigg[1+\frac1{c^2}\bigg(\frac{V^2}2+3U\bigg)\bigg].
\een
The notation for the mass density $\rho_*$ was introduced by Fock \cite{Fock} and the above equation corresponds to eq. \emph{(117)} of Chandrasekhar \cite{Ch1}.

In order to get the continuity equation in  the second post-Newtonian approximation  we have to transform the underlined terms in  (\ref{ba2a}) as follows. The first   underlined term can be rewritten as
\ben\lb{ba2f}
&&2\frac\rho{c^2}\bigg(\frac{\partial U }{\partial t}+V_i\frac{\partial U}{\partial x^i}\bigg)=\frac2{c^2}\bigg(\frac{\partial \rho U }{\partial t}+\frac{\partial \rho V_iU}{\partial x^i}\bigg)
+\frac{U}{c^4}\bigg[\bigg(\frac{\partial \rho V^2 }{\partial t}+\frac{\partial \rho V^2 V_i}{\partial x^i}\bigg)
+6\bigg(\frac{\partial \rho U }{\partial t}+\frac{\partial \rho U V_i}{\partial x^i}\bigg)\bigg],
\een
where the expression for the continuity equation in the first post-Newtonian approximation (\ref{ba2c}) was used. The second underlined term can be transformed according to
\ben\no
&&\frac\rho{c^4}\bigg(\frac{\partial U }{\partial t}+V_i\frac{\partial U}{\partial x^i}\bigg)\big(V^2-4U\big)=\frac1{c^4}\bigg\{\frac{\partial \rho UV^2 }{\partial t}+\frac{\partial \rho UV^2 V_i}{\partial x^i}-U\bigg(\frac{\partial \rho V^2 }{\partial t}+\frac{\partial \rho V^2 V_i}{\partial x^i}\bigg)
\\\lb{ba2g}
&&\qquad-4U\bigg[\frac{\partial \rho U }{\partial t}+\frac{\partial \rho U V_i}{\partial x^i}-U\underline{\bigg(\frac{\partial \rho  }{\partial t}+\frac{\partial \rho V_i}{\partial x^i}\bigg)}\bigg]\bigg\}.
\een
Here we note that for the above underlined term the Newtonian continuity equation (\ref{ba2c}) can be used  so that this term vanishes. Now by adding the two
equation (\ref{ba2f}) and (\ref{ba2g}) we get
\ben\no
&&2\frac\rho{c^2}\bigg(\frac{\partial U }{\partial t}+V_i\frac{\partial U}{\partial x^i}\bigg)+\frac\rho{c^4}\bigg(\frac{\partial U }{\partial t}+V_i\frac{\partial U}{\partial x^i}\bigg)\big(V^2-4U\big)=\frac2{c^2}\bigg(\frac{\partial \rho U }{\partial t}+\frac{\partial \rho UV_i}{\partial x^i}\bigg)+\frac1{c^4}\bigg\{\frac{\partial \rho UV^2 }{\partial t}+\frac{\partial \rho UV^2 V_i}{\partial x^i}
\\\lb{ba2h}
&&+2U\bigg(\frac{\partial \rho U }{\partial t}+\frac{\partial \rho UV_i}{\partial x^i}\bigg)\bigg\}=\frac2{c^2}\bigg(\frac{\partial \rho U }{\partial t}+\frac{\partial \rho UV_i}{\partial x^i}\bigg)+\frac1{c^4}\bigg\{\frac{\partial \rho UV^2 }{\partial t}+\frac{\partial \rho UV^2 V_i}{\partial x^i}+\bigg(\frac{\partial \rho U^2 }{\partial t}+\frac{\partial \rho U^2V_i}{\partial x^i}\bigg)\bigg\}+\mathcal{O}\left(c^{-6}\right).\qquad
\een
The last underlined term in (\ref{ba2a}) can be written as
\ben\no
-\frac\rho{c^4}\bigg[2\bigg(\frac{\partial \Phi }{\partial t}+V_i\frac{\partial \Phi}{\partial x^i}\bigg)+\frac12\bigg(\frac{\partial \Psi_{kk} }{\partial t}+V_i\frac{\partial \Psi_{kk}}{\partial x^i}\bigg)\bigg]=-\frac1{c^4}\Bigg[\frac{\partial \rho\left(2\Phi+\frac{\Psi_{kk}}2\right) }{\partial t}+\frac{\partial \rho V_i\left(2\Phi+\frac{\Psi_{kk}}2\right) }{\partial x^i}\Bigg]
\\\lb{ba2j}
+\frac1{c^4}\left(2\Phi+\frac{\Psi_{kk}}2\right)\underline{\left(\frac{\partial \rho }{\partial t}+\frac{\partial \rho V_i}{\partial x^i}\right)},\qquad
\een
where the underlined term above vanishes thanks to the Newtonian continuity equation (\ref{ba2c}).

The continuity equation in the second post-Newtonian approximation is obtained from (\ref{ba2a}) by using (\ref{ba2f})-- (\ref{ba2j}), yielding
\ben\lb{ba2k}
 \frac{\partial\widetilde\rho}{\partial t}+\frac{\partial \widetilde \rho V_i}{\partial x^i}=0,\qquad\hbox{where}\qquad \widetilde\rho=\rho\left[1+\frac1{c^2}\left(\frac{V^2}2+3U\right)+\frac1{c^4}\left(\frac38V^4+\frac72UV^2+\frac32U^2-\frac12\Psi_{kk}
 -\Pi_iV_i\right)\right].
 \een
The expression for $\widetilde\rho$ for the mass density in the second post-Newtonian approximation corresponds to eq. \emph{(53)} of  Chandrasekhar and Nutku \cite{ChNu}  and was determined from the consideration that the volume integral of $\rho U^0\sqrt{-g}$ is constant which is a consequence of the particle four-flow conservation equation.

\subsection{Mass-energy density hydrodynamic equation} 

The mass-energy density hydrodynamic equation is obtained by applying the same methodology, i.e. the  Boltzmann equation (\ref{k9}) is multiplied by $m^4u^0\sqrt{-g}d^3u/u_0$, the Maxwell-J\"uttner distribution function (\ref{mj4}) and  the integration element (\ref{mj8})  are used and the resulting equation integrated. The result is 
\ben\no
\frac{\partial}{\partial t}\bigg\{\rho\bigg[1+\frac1{c^2}\bigg(V^2+2U+\varepsilon\bigg)+\frac1{c^4}\bigg(V^4+6V^2U
+2U^2+V^2\bigg(\varepsilon+\frac{p}\rho\bigg)+2U\varepsilon+4\Phi-2\Pi_jV_j\bigg)\bigg]\bigg\}
\\\no
+\frac{\partial}{\partial x^i}\bigg\{\rho V_i\bigg[1+\frac1{c^2}\bigg(V^2+2U+\varepsilon+\frac{p}\rho\bigg)
+\frac1{c^4}\bigg(V^4+6V^2U+2U^2+4\Phi+(2U+V^2)\bigg(\varepsilon+\frac{p}\rho\bigg)
\\\no
-2\Pi_jV_j\bigg)\bigg]-\frac{p\Pi_i}{c^4}\bigg\}+\frac\rho{c^2}\frac{\partial U}{\partial t}+\frac\rho{c^4}\bigg[\bigg(\frac{3kT}{2m}+2U+2V^2\bigg)\frac{\partial U}{\partial t}-3\bigg(\frac{\partial U^2}{\partial t}+V_i\frac{\partial U^2}{\partial x^i}\bigg)+\Pi_i\frac{\partial U}{\partial x^i}
\\\lb{ba3a}
-\frac12\bigg(\frac{\partial \Psi_{kk}}{\partial t}+V_i\frac{\partial \Psi_{kk}}{\partial x^i}\bigg)
-\bigg(4\frac{\partial \Phi}{\partial t}+6V_i\frac{\partial \Phi}{\partial x^i}\bigg)+V_iV_j\frac{\partial \Pi_i}{\partial x^j}\bigg]=0.
\een
By neglecting all terms in $1/c^2$ and $1/c^4$ we get the Newtonian continuity equation (\ref{ba2b}). 

The first post-Newtonian approximation to the mass-energy hydrodynamic equation is obtained from (\ref{ba3a}) by neglecting the $1/c^4$ terms, yielding
\ben\lb{ba3b}
&&\frac{\partial\sigma}{\partial t}+\frac{\partial \sigma V_i}{\partial x^i}+\frac1{c^2}\left(\rho\frac{\partial U}{\partial t}-\frac{\partial p}{\partial t}\right)=0,\qquad\hbox{where}\qquad\sigma=\rho\bigg[1+\frac1{c^2}\bigg(V^2+2U+\varepsilon+\frac{p}\rho\bigg)\bigg].
\een
The abbreviation $\sigma$ was introduced by Chandrasekhar \cite{Ch1} and this  equation corresponds to the eq. \emph{(64)} of that work.

If we introduce the the abbreviations 
\ben\lb{ba3c}
&&\varphi=V^2+U+\frac\varepsilon2+\frac{3p}{2\rho}\\
&&\widetilde\sigma=\rho\bigg\{1+\frac1{c^2}\bigg(V^2+2U+\varepsilon+\frac{p}\rho\bigg)+\frac1{c^4}\bigg[V^4+6V^2U-U^2
+2U\varepsilon+V^2\bigg(\varepsilon+\frac{p}\rho\bigg)-\Pi_iV_i-\frac12\Psi_{kk}\bigg]\bigg\}.
\een
we can rewrite the second approximation to the mass-energy hydrodynamic equation   (\ref{ba3a}) in the following form
\ben\lb{ba3d}
\frac{\partial\widetilde\sigma}{\partial t}+\frac{\partial \widetilde\sigma V_i}{\partial x^i}+\frac1{c^2}\left(\rho\frac{\partial U}{\partial t}-\frac{\partial p}{\partial t}\right)+\frac{2\rho}{c^4}\bigg[\varphi\frac{\partial  U}{\partial t}-V_i\frac{\partial\Phi}{\partial x^i}
+\frac1\rho\frac{\partial pUV_i}{\partial x^i}-\frac{V_i}2\frac{\partial \Pi_i}{\partial t}\bigg]=0.
\een
In this equation the term
\ben\lb{ba3e}
\frac1{c^4}\left(3U^2+4\Phi+\frac{\Psi_{kk}}2\right)\left[\frac{\partial\rho}{\partial t}+\frac{\partial\rho V_i}{\partial x^i}\right]
-\frac{\Pi_i}{c^4}\left[\frac{\partial\rho V_i}{\partial t}+\frac{\partial\rho V_iV_j}{\partial x^j}+\frac{\partial p}{\partial x^i}-\rho\frac{\partial U}{\partial x^i}\right].
\een
was neglected, since the Newtonian continuity equation (\ref{ba2b}) and the momentum hydrodynamic equation (\ref{ba4b}) for the first and the second terms within the brackets above can be used, respectively.

Equation (\ref{ba3d}) has not been derived in the work of Chandrasekhar and Nutku \cite{ChNu} but we shall see that it is important to determine the post-Newtonian hydrodynamic equation for the internal energy density.

\subsection{Momentum density hydrodynamic equation}
The hydrodynamic equation for the momentum density is obtained from the multiplication of Boltzmann equation (\ref{k9})   by $m^4u^i\sqrt{-g}d^3u/u_0$ and integration of the resulting equation, taking into account  the Maxwell-J\"uttner distribution function (\ref{mj4}) and  the integration element (\ref{mj8}). Up to   the first post-Newtonian approximation we get 
\ben\no
\frac{\partial}{\partial t}\bigg\{\rho\bigg[1+\frac1{c^2}\bigg(V^2+2U+\varepsilon+\frac{p}\rho\bigg)\bigg]V_i\bigg\}+\frac{\partial}{\partial x^j}\bigg\{\rho V_iV_j\bigg[1+\frac1{c^2}\bigg(V^2+2U+\varepsilon+\frac{p}\rho\bigg)\bigg]\bigg\}+\frac{\partial}{\partial x^i}\bigg[p\bigg(1-\frac{2U}{c^2}\bigg)\bigg]\qquad
\\\lb{ba4a}
-\rho\bigg[1+\frac1{c^2}\bigg(2V^2-2U+\varepsilon-\frac{p}\rho\bigg)\bigg]\frac{\partial U}{\partial x^i}+4\frac\rho{c^2}V_i\bigg(\frac{\partial U}{\partial t}+V_j\frac{\partial U}{\partial x^j}\bigg)
-\frac{\rho}{c^2}\bigg[\frac{\partial\Pi_i}{\partial t}+V_j\bigg(\frac{\partial\Pi_i}{\partial x^j}-\frac{\partial\Pi_j}{\partial x^i}\bigg)+2\frac{\partial\Phi}{\partial x^i}\bigg]
=0.\qquad
\een
By neglecting the $1/c^2$ terms we get the Newtonian momentum density hydrodynamic equation 
\ben\lb{ba4b}
\frac{\partial\rho V_i}{\partial t}+\frac{\partial\rho V_iV_j}{\partial x^j}+\frac{\partial p}{\partial x^i}
-\rho\frac{\partial U}{\partial x^i}=0.
\een
The following terms can be rewritten in the equivalent forms
\ben\lb{ba4c}
&&\frac{\partial}{\partial x^i}\left[p\left(1-\frac{2U}{c^2}\right)\right]=\frac{\partial p}{\partial x^i}\left(1-\frac{2U}{c^2}\right)-\frac{2p}{c^2}\frac{\partial U}{\partial x^i},
\\\no
&&4\frac\rho{c^2}V_i\bigg(\frac{\partial U}{\partial t}+V_j\frac{\partial U}{\partial x^j}\bigg)=\frac4{c^2}\left(\frac{\partial \rho UV_i}{\partial t}+\frac{\partial \rho UV_iV_j}{\partial x^j}\right)-\frac4{c^2}U\left(\frac{\partial \rho V_i}{\partial t}+\frac{\partial \rho V_iV_j}{\partial x^j}\right)
\\\lb{ba4d}
&&\qquad
=\frac4{c^2}\left(\frac{\partial \rho UV_i}{\partial t}+\frac{\partial \rho UV_iV_j}{\partial x^j}\right)-\frac4{c^2}U\left(\frac{\partial p}{\partial x^i}-\frac{\partial U}{\partial x^i}\right)+\mathcal{O}(c^{-4}),
\\\lb{ba4e}
&&\frac{\rho}{c^2}\left(\frac{\partial\Pi_i}{\partial t}+V_j\frac{\partial\Pi_i}{\partial x^j}\right)=\frac{1}{c^2}\left(\frac{\partial\rho\Pi_i}{\partial t}+\frac{\partial\rho\Pi_i V_j}{\partial x^j}\right)-\frac{\Pi_i}{c^2}\left(\frac{\partial\rho}{\partial t}+\frac{\partial\rho V_j}{\partial x^j}\right)=\frac{1}{c^2}\left(\frac{\partial\rho\Pi_i}{\partial t}+\frac{\partial\rho\Pi_i V_j}{\partial x^j}\right)+\mathcal{O}(c^{-4}),
\een
where the Newtonian hydrodynamic equations for the momentum density (\ref{ba4b}) and for the mass density  (\ref{ba2b}) were used in the equations (\ref{ba4d}) and (\ref{ba4e}), respectively.
By collecting the above results the momentum density hydrodynamic equation (\ref{ba4a}) in the first post-Newtonian approximation can be rewritten as 
\ben\lb{ba4f}
\frac{\partial\rho\frak{V}_i}{\partial t}+\frac{\partial\rho\frak{V}_iV_j}{\partial x^j}+\frac{\partial p}{\partial x^i}\left[1+\frac{2U}{c^2}\right]
-\rho\frac{\partial U}{\partial x^i}\bigg[1+\frac2{c^2}\bigg(V^2+U+\frac\varepsilon2+\frac{3p}{2\rho}\bigg)\bigg]
+\frac\rho{c^2}\bigg(V_j\frac{\partial \Pi_j}{\partial x^i}-2\frac{\partial \Phi}{\partial x^i}\bigg)
=0.
\een
Here we have introduced the following abbreviation for the momentum density 
\ben\lb{ba4g}
\rho\frak{V}_i=\rho V_i\left[1+\frac1{c^2}\left(V^2+6U+\varepsilon+\frac{p}\rho\right)\right]-\frac\rho{c^2}\Pi_i.
\een

A more familiar equation for the momentum density hydrodynamic equation (\ref{ba4f})  is obtained when the material time derivative $d/dt=\partial/\partial t+V_i\partial/\partial x^i$ is introduced. After some rearrangements this equation becomes
\ben\no
\rho\frac{dV_i}{dt}+\frac{\partial p}{\partial x^i}\left[1-\frac1{c^2}\left(V^2+4U+\varepsilon+\frac{p}\rho\right)\right]-\rho\frac{\partial U}{\partial x^i}\left[1+\frac1{c^2}\left(V^2-4U\right)\right]
\\\lb{ba41}
-\frac\rho{c^2}\left[2\frac{\partial \Phi}{\partial x^i}+\frac{d\Pi_i}{dt}-V_j\frac{\partial \Pi_j}{\partial x^i}+V_i\left(\frac{\partial U}{\partial t}-\frac1\rho\frac{\partial p}{\partial t}-4\frac{dU}{dt}\right)\right]=0.
\een

As was pointed in \cite{ChNu} the hydrodynamic equation for the momentum density in the second post-Newtonian approximation  follows  after a long calculation, yielding
\ben\no
&&\frac{\partial \rho\widetilde{\frak{V}}_i}{\partial t}+\frac{\partial \rho\widetilde{\frak{V}}_iV_j}{\partial x^j}+\frac{\partial p}{\partial x^i}\left[1+\frac{2U}{c^2}-\frac1{c^4}\left(U^2+2\Phi+\frac{\Psi_{kk}}2\right)\right]
-\rho\frac{\partial U}{\partial x^i}\bigg\{1+\frac2{c^2}\bigg(V^2+U+\frac\varepsilon2+\frac{3p}{2\rho}\bigg)+\frac2{c^4}\bigg[V^4+5UV^2
\\\no
&&\qquad-\frac{3U^2}2+\Phi+(V^2+U)\bigg(\varepsilon+\frac{p}\rho\bigg)-\Pi_iV_i-\frac{\Psi_{kk}}4\bigg]\bigg\}
+\frac\rho{c^2}\bigg(V_j\frac{\partial \Pi_j}{\partial x^i}-2\frac{\partial \Phi}{\partial x^i}\bigg)\bigg[1+\frac1{c^2}\bigg(V^2+4U+\varepsilon+\frac{p}\rho\bigg)\bigg]
\\\lb{ba4h}
&&\qquad+\frac\rho{2c^4}\bigg(\frac{\partial \Psi_{00}}{\partial x^i}+2V_j\frac{\partial \Psi_{0j}}{\partial x^i}+V_jV_k\frac{\partial \Psi_{jk}}{\partial x^i}\bigg)=0,
\een
where the abbreviation for the momentum density was introduced
\ben\no
&&\rho\widetilde{\frak{V}}_i=\rho V_i\bigg\{1+\frac1{c^2}\bigg(V^2+6U+\varepsilon+\frac{p}\rho\bigg)+\frac1{c^4}\bigg[V^4+10V^2U
+13U^2+2\Phi-2\Pi_iV_i-\frac{\Psi_{kk}}2
\\\lb{ba4i}
&&\quad+\big(V^2+6U\big)\bigg(\varepsilon+\frac{p}\rho\bigg)\bigg]\bigg\}-\frac\rho{c^2}\Pi_i\bigg[1+\frac1{c^2}\bigg(V^2+4U+\varepsilon+\frac{p}\rho\bigg)\bigg]-\frac\rho{c^4}\big(\Psi_{0i}+\Psi_{ij}V_j\big).
\een

\subsection{Total energy density hydrodynamic equation}

The total energy density of the gas is the sum of its internal   $\rho\varepsilon$ and  kinetic $\rho V^2/2$  energy densities. Its hydrodynamic equation can be obtained by subtracting the continuity equation (\ref{ba2b}) from the mass-energy hydrodynamic equation (\ref{ba3d}), yielding
\ben\no
&&\frac1{c^2}\bigg\{\frac{\partial}{\partial t}\bigg[{\rho\bigg(\frac{V^2}2+\varepsilon\bigg)+\frac\rho{c^2}\bigg[\frac58V^4+\frac52V^2U-\frac52U^2+2\varepsilon U+V^2\bigg(\varepsilon+\frac{p}\rho\bigg)\bigg]\bigg]}
\\\no
&&+\frac{\partial}{\partial x^i}\bigg[\rho V_i\bigg(\frac{V^2}2+\varepsilon\bigg)+\frac{\rho V_i}{c^2}\bigg[\frac58V^4+\frac52V^2U-\frac52U^2+2\varepsilon U+V^2\bigg(\varepsilon+\frac{p}\rho\bigg)\bigg]\bigg]
\\\lb{ba5a}
&&+\frac{\partial pV_i}{\partial x^i}-\rho V_i\frac{\partial U}{\partial x^i}-U\left(\frac{\partial\rho}{\partial t}+\frac{\partial\rho V_i}{\partial x^i}\right)+\frac{2\rho}{c^2}\bigg[\varphi\frac{\partial  U}{\partial t}-V_i\frac{\partial\Phi}{\partial x^i}+\frac1\rho\frac{\partial pUV_i}{\partial x^i}-\frac{V_i}2\frac{\partial \Pi_i}{\partial t}\bigg]\bigg\}=0.
\een
Note that  the total energy density hydrodynamic equation  is  of order $\mathcal{O}\left(c^{-2}\right)$, meaning that the post-Newtonian corrections to the resulting equation corresponds to the first post-Newtonian approximation.

 The hydrodynamic equation for the internal energy density in the first post-Newtonian approximation is obtained from  (\ref{ba5a}) by eliminating  the time derivative of the mass density $\rho$ and hydrodynamic velocity $V_i$ taking into account the first post-Newtonian   hydrodynamic equations for the mass density (\ref{ba2e}) and momentum density (\ref{ba4f}). After some rearrangements  we get that
 \ben
 \rho\frac{d\varepsilon}{dt}+p\left(1-\frac53\frac{V^2}{c^2}\right)\frac{\partial V_i}{\partial x^i}+\frac{\rho V_i}{ c^2}\left(V^2+\frac23\varepsilon\right)\left[\frac{\partial U}{\partial x^i}-\frac1\rho\frac{\partial p}{\partial x^i}\right]-\frac{V^2}{c^2}\frac{\partial p}{\partial t}
 +\frac{\rho\left(2\varepsilon-5U\right)}{c^2}\frac{dU}{dt}=0.
 \een
   Without the terms in $c^{-2}$ this equation reduces to the Newtonian internal energy density hydrodynamic equation for an Eulerian fluid, namely
 \ben
 \rho\frac{d\varepsilon}{dt}+p\frac{\partial V_i}{\partial x^i}=0.
 \een

\section{Conclusions}\lb{s.6}
In this work a kinetic theory of relativistic gases in the presence of gravitational fields within the second post-Newtonian approximation was developed. The Boltzmann equation, the equilibrium Maxwell-J\"uttner distribution function and the hydrodynamic equations were found in the same approximation. The expressions for the components of the particle four-flow, energy-momentum tensor and the hydrodynamic equations for the mass, mass-energy and momentum densities correspond to those obtained from a phenomenology theory.

\section*{Appendix A}
In this appendix we give the components of the Christoffel symbols corresponding to the components of the metric tensor (\ref{sp1a}) -- (\ref{sp1c}). 
\ben\lb{spp5a}
&&{\buildrel\!\!\!\! _{3} \over{\Gamma^0}_{00}}=-\frac1{c^3}\frac{\partial U}{\partial t},\qquad {\buildrel\!\!\!\! _{5} \over{\Gamma^0}_{00}}=\frac1{c^5}\left(\Pi_i\frac{\partial U}{\partial x^i}-2\frac{\partial \Phi}{\partial t}\right),\qquad
{\buildrel\!\!\!\! _{2} \over{\Gamma^0}_{0i}}=-\frac1{c^2}\frac{\partial U}{\partial x^i},
\qquad
{\buildrel\!\!\!\! _{4} \over{\Gamma^0}_{0i}}=-\frac2{c^4}\frac{\partial\Phi}{\partial x^i},
\\\lb{spp5b}
&&{\buildrel\!\!\!\! _{2} \over{\Gamma^i}_{00}}=-\frac1{c^2}\frac{\partial U}{\partial x^i},
\qquad{\buildrel\!\!\!\! _{4} \over{\Gamma^i}_{00}}=\frac2{c^4}\frac{\partial (U^2-\Phi)}{\partial x^i}-\frac1{c^4}\frac{\partial\Pi_i}{\partial t},
\qquad{\buildrel\!\!\!\! _{3} \over{\Gamma^0}_{ij}}=\frac1{2c^3}\left(\frac{\partial\Pi_i}{\partial x^j}+\frac{\partial\Pi_j}{\partial x^i}+2\frac{\partial U}{\partial t}\delta_{ij}\right),
\\\lb{spp5c}
&&{\buildrel\!\!\!\! _{2} \over{\Gamma^i}_{jk}}=\frac1{c^2}\left(\frac{\partial U}{\partial x^j}\delta_{ik}+\frac{\partial U}{\partial x^k}\delta_{ij}-\frac{\partial U}{\partial x^i}\delta_{jk}\right),
\qquad {\buildrel\!\!\!\! _{3} \over{\Gamma^j}_{0i}}=\frac1{2c^3}\left(\frac{\partial\Pi_i}{\partial x^j}-\frac{\partial\Pi_j}{\partial x^i}+2\frac{\partial U}{\partial t}\delta_{ij}\right).
\\
&&{\buildrel\!\!\!\! _{4} \over{\Gamma^i}_{jk}}=-\frac1{2c^4}\left(\frac{\partial\Psi_{ij}}{\partial x^k}+\frac{\partial\Psi_{ik}}{\partial x^j}-\frac{\partial\Psi_{jk}}{\partial x^i}\right)
-\frac1{c^4}\left(\frac{\partial U^2}{\partial x^k}\delta_{ij}+\frac{\partial U^2}{\partial x^j}\delta_{ik}-\frac{\partial U^2}{\partial x^i}\delta_{jk}\right).
\\
&&{\buildrel\!\!\!\! _{5} \over{\Gamma^i}_{0j}}=-\frac1{c^5}\Pi_i\frac{\partial U}{\partial x^j}+\frac{U}{c^5}\left(\frac{\partial\Pi_i}{\partial x^j}-\frac{\partial \Pi_j}{\partial x^i}-2\frac{\partial U}{\partial t}\delta_{ij}\right)
-\frac1{2c^5}\left(\frac{\partial \Psi_{0i}}{\partial x^j}-\frac{\partial \Psi_{0j}}{\partial x^i}+\frac{\partial\Psi_{ij}}{\partial t}\right).
\\
&&{\buildrel\!\!\!\! _{6} \over{\Gamma^i}_{00}}=\frac1{2c^6}\frac{\partial\Psi_{00}}{\partial x^i}-\frac1{c^6}\frac{\partial\Psi_{0i}}{\partial t}-\frac{\Pi_i}{c^6}\frac{\partial U}{\partial t}-\frac{4U}{c^6}\bigg(\frac{\partial U^2}{\partial x^i}-\frac{\partial \Phi}{\partial x^i}\bigg)
+\frac{2U}{c^6}\frac{\partial \Pi_i}{\partial t}-\frac{\Psi_{ij}}{c^6}\frac{\partial U}{\partial x^j}.
\een

\section*{Appendix \textbf{B}}
For the integration of the equations in Section \ref{s.4} we have used the  following expressions from the kinetic theory of gases (see e.g.\cite{GK})
\ben
&&I_n=\int \mathcal{V}^ne^{- \frac{m\mathcal{V}^2}{kT}}d\mathcal{V}=\frac12\Gamma\left(\frac{n+1}2\right)\left(\frac{kT}m\right)^\frac{n+1}2
\qquad
\Gamma(n+1)=n\Gamma(n),\qquad \Gamma(1)=1, \qquad \Gamma\left(\frac12\right)=\sqrt\pi,
\\
&&\int e^{- \frac{m\mathcal{V}^2}{kT_0}}\mathcal{V}_i\mathcal{V}_jd^3\mathcal{V}=\frac{I_{2}}{3} \delta_{ij},
\qquad
\int e^{- \frac{m\mathcal{V}^2}{kT_0}}\mathcal{V}_i\mathcal{V}_j\mathcal{V}_k\mathcal{V}_ld^3\mathcal{V}=\frac{I_{4}}{15} \left[\delta_{ij}\delta_{kl}+ \delta_{ik}\delta_{jl}+\delta_{il}\delta_{jk}\right],
\\\no
&&\int e^{- \frac{m\mathcal{V}^2}{kT_0}}\mathcal{V}_i\mathcal{V}_j\mathcal{V}_k\mathcal{V}_l\mathcal{V}_n\mathcal{V}_nd^3\mathcal{V}
=\frac{I_{6}}{105}\big[\delta_{ij}\big(\delta_{kl}\delta_{mn} + \delta_{km}\delta_{ln}+\delta_{kn}\delta_{lm}\big)+ \delta_{ik}\big(\delta_{jl}\delta_{mn} +\delta_{jm}\delta_{ln} + \delta_{jn}\delta_{lm}\big)
\\
&&+\delta_{il}\big(\delta_{jk}\delta_{mn}+ \delta_{jm}\delta_{kn}+\delta_{jn}\delta_{km}\big)
 + \delta_{im}\big(\delta_{jk}\delta_{ln}+\delta_{jl}\delta_{kn}+ \delta_{jn}\delta_{kl}\big)+
\delta_{in}\big(\delta_{jk}\delta_{lm}+\delta_{jl}\delta_{km}+\delta_{jm}\delta_{kl}\big)\big].
\een

\end{document}